\documentclass[11pt]{article}
\usepackage{natbib}
\usepackage[english]{babel} 
\usepackage{graphicx}
\usepackage{times,amssymb}
\usepackage[fleqn,tbtags]{amsmath}
\usepackage[varg]{txfonts}
\usepackage{amsmath}
\usepackage{txfonts}

\usepackage{rotating}
\usepackage{dcolumn}

\usepackage[utf8x]{inputenc}
\usepackage{amssymb}
\usepackage{amsfonts}
\usepackage{array}
\usepackage{multirow}
\usepackage{etoolbox}
\usepackage{amstext}
\usepackage{amsxtra}
\usepackage{xcolor}
\usepackage{footnote}

\def\lsim{~\rlap{$<$}{\lower 1.0ex\hbox{$\sim$}}}

\def\gsim{~\rlap{$>$}{\lower 1.0ex\hbox{$\sim$}}}

\begin{document}

\title{Prospects of detecting HI using redshifted $21$~cm radiation at $z \sim 3$}   

\author{Bharat Kumar Gehlot$^{1}$ and J.S. Bagla$^2$ \\
$^1$Kapteyn Astronomical Institute,  University of Groningen, Netherlands \\
  $^2$Department of Physical Sciences,  \\
  Indian Institute of Science Education and Research Mohali, Sector 81, S.A.S. Nagar, Punjab 140306, India \\
E-Mail: gehlot@astro.rug.nl, jasjeet@iisermohali.ac.in}

\maketitle

\begin{abstract}
Distribution of cold gas in the post-reionization era provides an
important link between distribution of galaxies and the process of
star formation.  
Redshifted $21$~cm radiation from the Hyperfine transition of neutral
Hydrogen allows us to probe the neutral component of cold gas, most of
which is to be found in the interstellar medium of galaxies.
Existing and upcoming radio telescopes can probe the large scale
distribution of neutral Hydrogen via HI intensity mapping. 
In this paper we use an estimate of the HI power spectrum derived
using an ansatz to compute the expected signal from the large scale HI
distribution at $z \sim 3$. 
We find that the scale dependence of bias at small scales makes a
significant difference to the expected signal even at large angular
scales.
We compare the predicted signal strength with the sensitivity of
radio telescopes that can observe such radiation and calculate the
observation time required for detecting neutral Hydrogen at these
redshifts. 
We find that OWFA (Ooty Wide Field Array) offers the best possibility 
to detect neutral Hydrogen at $z \sim 3$ before the SKA (Square
Kilometer Array) becomes operational. 
We find that the OWFA should be able to make a $3\, \sigma$ or
a more significant detection in $2000$~hours of observations at
several angular scales.
Calculations done using the Fisher matrix approach indicate that
a $5\sigma$ detection of the binned HI power spectrum via measurement
of the amplitude of the HI power spectrum is possible in $1000$~hours
\citep{OWFA.HIPK}. 
\end{abstract}

\bigskip


\bigskip

\section{Introduction}

It is believed that the formation of galaxies in dark matter halos led
to emission of UV radiation that gradually ionized the inter-galactic
medium (IGM) by $z \simeq 6$ \citep{2013ASSL..396...45Z}. 
The IGM is almost fully ionized at lower redshifts, and nearly all the
neutral Hydrogen in the universe is to be found in the inter-stellar
medium (ISM) of galaxies \citep{2005ARA&A..43..861W,2013ARA&A..51..105C}.
This allows us to model the distribution of neutral gas in terms of 
assignment schemes that apportion the neutral gas in halos of dark
matter.  
Such a model can then be constrained by observations of damped Lyman
$\alpha$ systems (DLAS) seen as strong absorption in the spectra of 
quasars at high redshifts \citep{2014JCAP...09..050V,2015MNRAS.447.3745P}.
The observations of DLAS indicate that at redshifts $2 \leq z
\leq 5$, the neutral hydrogen content of the universe is almost
constant with a density parameter ${\Omega}_{HI} \sim 0.001$
\citep{2012A&A...547L...1N}.
Clustering of DLAS has been used to constrain the {\sl bias} for these
objects \citep{2012JCAP...11..059F} and it appears that these objects
are strongly clustered when compared with the matter distribution at
these redshifts.

Several upcoming radio telescopes can be used to detect neutral
Hydrogen at high redshifts using the redshifted $21$~cm radiation
using intensity mapping
\citep{2001JApA...22..293B,2015aska.confE..19S,2013PASA...30....7T}. 
The focus of the present study is at redshift $z \simeq 3$.
A number of telescopes can potentially detect redshifted $21$~cm from
neutral Hydrogen at high redshifts:  
GMRT (Giant Meterwave Radio
Telescope)\footnote{http://www.gmrt.ncra.tifr.res.in}, uGMRT
(upgraded 
GMRT), OWFA (Ooty Wide Field Array), MWA (Murchison Widefield
Array)\footnote{http://www.mwatelescope.org/}, CHIME (Canadian
Hydrogen Intensity Mapping
Experiment)\footnote{http://chime.phas.ubc.ca} and
SKA-low\footnote{https://www.skatelescope.org/}.
The focus of the present study is $z \sim 3$.  
Thus we discuss the feasibility of detecting redshifted $21$~cm
radiation using the GMRT, MWA, SKA and OWFA as these instruments can
target this redshift range.
The analysis is not meant to be exhaustive but a broad brush
comparison of these instruments. 

In this paper we use the model suggested by \citet{2010MNRAS.407..567B} for populating dark matter halos with neutral Hydrogen.   
We then proceed to compute the expected signal and compare this with
the sensitivity of several radio telescopes.  
This comparison is done using the approach used in \citep{2014JApA...35..157A}. 
We do not consider the foregrounds in this paper, readers may refer to
\citet{2014JApA...35..157A} for a discussion and references.

\section{Methodology}

The observables in the context of the HI intensity mapping are
visibility correlations and these have been shown to be related to the
redshift space power spectrum of HI fluctuations
\citep{2001JApA...22..293B,2008PhRvL.100i1303C} in the limit of a
small field of view.    
A physically motivated scheme to populate dark matter halos with
neutral Hydrogen has been used in the past for modelling the HI power
spectrum \citep{2010MNRAS.407..567B} and similar results for
clustering are obtained in the halo model with a similar approach
\citep{2010MNRAS.404..876W}.  
This model is able to reproduce the column density distribution of
DLAS \citep{2014JCAP...09..050V,2015MNRAS.447.3745P}, 
though it seems to under-predict the bias for DLAS as compared to
recent observational determination \citep{2012JCAP...11..059F}. 

In this paper we use a fitting function for the HI power spectrum in
order to estimate the signal in redshifted $21$~cm radiation from the
large scale distribution of HI at $z \sim 3$.
This approach uses the linearly evolved matter power spectrum and a
fitting function for a scale dependent bias and non-linear evolution,
where the fitting function is obtained from N-Body simulations
\citep{2010MNRAS.407..567B}.   
Figure~1 shows the power spectrum for the HI distribution as obtained
from N-Body simulations.  
For ease of computation, we do not use the conventional bias but work
instead with the offset between the HI power spectrum and the linearly
extrapolated power spectrum for dark matter: thus it gives the
combined effect of non-linear growth of perturbations as well as a
scale dependent bias. 
One immediate benefit of this approach is that we are not limited to
the scales resolved in simulations used to derive the HI power
spectrum. 
There is a need to excercise caution when extrapolating to smaller scales
as we know little about the scale dependence of bias at these scales
from N-Body simulations. 
However, in this work we are analyzing signal primarily at large
angular scales and hence the effect of errors in modelling scales
smaller than those resolved in simulations can be shown to be
minimal.

The HI content of galaxies at very low redshifts can be estimated
directly from observed emission of $21$~cm radiation
\citep{2005MNRAS.359L..30Z,2016A&ARv..24....1G}.
This radiation arises from the Hyperfine transition of neutral
Hydrogen \citep{1952ApJ...115..206W,1957BAN....14....1V} and
if the spin temperature is much higher than the temperature of the
CMBR (cosmic microwave background radiation) then the emission is
proportional to the amount of neutral Hydrogen
\citep{2006PhR...433..181F}.
The spin temperature in turn is determined by coupling with atoms,
electrons and Lyman-$\alpha$ line radiation
\citep{1952AJ.....57R..31W,1952Phy....18...75W,1956ApJ...124..542P,
  2006PhR...433..181F}. 
At intermediate redshifts, techniques such as co-adding
\citep{2001A&A...372..768C,2000PhDT..........Z,2007MNRAS.376.1357L,
  2013MNRAS.435.2693R,2013MNRAS.433.1398D,2016MNRAS.460.2675R,
  2016ApJ...818L..28K} 
and cross-correlation \citep{2010Natur.466..463C,2013ApJ...763L..20M}
have been used to detect neutral Hydrogen.
As we get to even higher redshifts, statistical detection is the only
viable approach for observing the large scale distribution of HI. 
The collective emission from the undetected regions is present as a
very faint background in all radio observations at frequencies below
$1420$~MHz.  
The fluctuations in this background radiation carry an imprint of the
HI distribution at the redshift $z$ where the radiation originated.
HI emission from the post-reionization era can be used as a probe of
the large scale structure and can also be used to constrain
cosmological paramters \citep{2001JApA...22..293B,2009PhRvD..79h3538B,
  2012RPPh...75h6901P}.
It should be noted though that the possibility of direct detection of
extreme objects \citep{1997MNRAS.289..671B} should not be
discounted, particularly in the light of very high bias at small
scales \citep{2010MNRAS.407..567B,2010MNRAS.404..876W}.

Realizing the potential for constraining cosmological parameters, the
majority of the existing and upcoming radio-interferometric
experiments such as LOFAR, MWA, SKA, PAPER, etc. are aimed at
measuring the HI 21 cm signal statistically and map out the large
scale HI distribution at high redshifts with the primary focus on the
epoch of reionization.  
Similarly CHIME is an experiment for detecting post-reionization
distribution of HI at large scales. 
The Ooty Wide Field Array (OWFA) \citep{2011ASInC...3..165B,OWFA2017}
is the upgraded Ooty radio telescope where the
collecting area and the primary receivers remain the same but the
entire backend has been changed. 
OWFA has a much larger field of view and a much larger bandwidth as
compared to the earlier configuration.
A new programmable backend has been added to the existing capability. 
For details, please see
\citep{1971NPhS..230..185S,1975JIETE..21..117K,1975JIETE..21..110S,
  1988BASI...16..111J,2011ExA....31....1P,OWFA2014,OWFA2017}.
OWFA has the potential of detecting HI at $z \simeq 3$ and also
measuring the power spectrum \citep{2014JApA...35..157A}. 

\section{Ooty Wide Field Array}

OWFA consists of a parabolic cylindrical antenna which is $530$~m long in
north-south direction and $30$~m wide in east-west direction.
It is an equatorially mounted phased array of $1056$ dipoles placed
along the focal line of the reflector.
The East-West steering (RA) is done by the mechanical rotation of the
telescope about its N-S axis and steering of beam in north-south
direction (Declination) is achieved by introducing appropriate delays
between the dipoles.
The array possesses huge redundancy (different antenna pairs
correspond to same baseline) due to its linearity and uniform spacing
of the antenna elements \citep{2011ExA....31....1P}.
This redundancy is very useful for precision experiments, since it
allows for model independent estimation of both the element gains as
well as true visibilities \citep{1991ASPC...19..192W,
  1992ExA.....2..203W,2010MNRAS.408.1029L,2014MNRAS.437..524M}.
The array has been upgraded to improve sensitivity and the Field of
View (FoV). 
It can now be used as an interferometer and has multibeaming feature.  

OWFA has $264$ antenna elements and each element can be
approximated as a rectangular aperture dimensions(\textsl{b $\times$
  d}) equal to $30$~m $\times$ $1.92$~m.
The primary field of view is about $1.75$\textdegree \ in E-W
direction and $27.4$\textdegree \ in N-S direction with a resolution
of $0.1$\textdegree .
The smallest baseline corresponds to antenna separation of $1.92$~m and the
largest baseline is $505$~m.
Total bandwidth of the system is $39$~MHz. 

The upgraded array is sensitive primarily to HI power spectrum at
wave numbers between $0.02 - 0.2$~h$/$Mpc \citep{2015JApA...36..385B,
  OWFA.HIPK}. 
The observable signal here is the visibility correlation and it is
related to an integral over the power spectrum and hence it has some
contribution from power spectrum at other scales
\citep{2001JApA...22..293B, 2005MNRAS.356.1519B}.

If we compare OWFA with the other radio telescopes that are, or will
be, sensitive to redshifted $21$~cm radiation from $z \sim 3$ then we
find that each has its own peculiarities.
\begin{itemize}
\item
  GMRT has a very large collecting area and hence very good
  sensitivity.  However, the primary beam is very small as the size of
  the dish is large at $45$~m.
\item
  MWA has a small effective collecting area but the field of view
  is large and it is sensitive to signal at a wide range of angular
  scales.
\item
  SKA-Low is in some sense akin to MWA but with a much larger
  collecting area.  
\end{itemize}

\begin{table}
\begin{center}
\caption{System parameters for OWFA}
\begin{tabular}{| p{9.5cm} |  >{\centering\arraybackslash}m{3.5cm} |}	
	\hline 
	{\bf Parameter} &  {\bf OWFA} \\ \hline
	No. of antennas ($N_A$) & 264 \\ \hline
	Aperture dimensions ($b \times d$) & $30$ m $\times$ $1.92$ m
        \\
        \hline
	Field of View(FoV) & $1.75$\textdegree $\times$
        $27.4$\textdegree \\
        \hline 
	Smallest baseline ($d_{min}$) & $1.92$ m \\ \hline
	Largest baseline ($d_{max}$) & $505.0$ m	\\ \hline
	Angular Resolution & $1.75^\circ \times 6.3'$ \\ \hline	
	Total Bandwidth (B) & $30$ MHz \\ \hline
	Single Visibility rms. noise ($\sigma$) assuming $T_{sys =}$
        150 K, $\eta =$ 0.6, $\Delta \nu_c =$ 0.1 MHz, $\Delta t =$ 16
        s & \multirow{3}*{6.69 Jy} \\ \hline  
\end{tabular}
\end{center}
\end{table}

\section{Visibility correlations and HI Power Spectrum}

The multi-frequency angular power spectrum $C_l(\Delta \nu)$ can be
used to quantify the statistical properties of the signal observed on
the sky.
It is a function of the angular scale $l$ and frequency shift
$\Delta\nu$. 
The observed visibilities $ {\cal V}(U_n,\Delta \nu)$
are related to $C_l(\Delta \nu)$ through the visibility
correlation which can be written as
\citep{2007MNRAS.378..119D}
\begin{equation}
C_l(\Delta \nu) = 0.26
\ {\left(\dfrac{\text{mK}}{\text{Jy}}\right)}^{2}
\left(\dfrac{bd}{m^2}\right) V_2(U_n, \Delta \nu)  
\end{equation}
Where 
\begin{equation}
V_2(U_n, \Delta \nu) \equiv \langle{\cal V} (U_n,\nu) {\cal V}^*
(U_m,\nu + \Delta \nu)\rangle 
\end{equation} 
is the visibility correlation.

Equation (1) provides a reasonably good approximation for the entire
baseline range covered by OWFA.
The measured visibility correlations are a combination of three
different components:
\begin{equation}
{V}_{2}(U_n,\Delta \nu) = {S}_{2}(U_n,\Delta \nu) + {F}_{2}(U_n,\Delta
\nu) + {N}_{2}(U_n,\Delta \nu)  
\end{equation}
where $S_2, F_2$ and $N_2$ respectively are the HI signal, foreground
and the noise contribution to the visibility correlation.
In our calculations we have assumed that all foreground signals are
already removed from the signal (zero foreground contribution) and
treat the total signal as consisting of only HI signal and noise.
Removal of foregrounds is an essential and a non-trivial task: see,
e.g., \citep{2008MNRAS.385.2166A, 2011MNRAS.418.2584G, 2015MNRAS.452.1587G}.
A discussion of foreground removal is beyond the scope of this
paper.  

${S}_{2}(U_n,\Delta \nu)$ from the HI signal directly
probes the redshift space power spectrum $P_{HI}(\textbf{k},z)$
\citep{2001JApA...22..293B,2005MNRAS.356.1519B}. 
The multi-frequency angular power spectrum can be calculated for the 
HI signal using equation (4).
\begin{equation}
C_l(\Delta \nu) = \dfrac{1}{\pi r_{\nu}^2} \int_0^{\infty} dk_{||}
\text{cos} (k_{||}r_{\nu}' \Delta \nu) P_{\text{HI}}(\textbf{k}) 
\end{equation}
where \textbf{k} has magnitude $k = \sqrt{k_{\parallel}^{2} +
  l^2/r_{\nu}^2}$ and has components $k_{\parallel}$ and $l/r_{\nu}$
along the line of sight and in the plane of the sky respectively.  The
multipole index $l$ takes on positive integer values.  
Here $r_{\nu}$ is the comoving distance corresponding to $z =
(1420\text{MHz}/\nu) - 1$, $r_{\nu}' = \frac{dr_{\nu}}{d\nu}$. 

Assuming that HI traces the total matter distribution with a bias
parameter, $P_{HI}(k,\mu) \equiv P_{HI}(\textbf{k})$ can be modelled
as  
\begin{equation}
P_{\text{HI}} (k,\mu) = b^2 \bar{x}_{\text{HI}}^{2}\bar{T}^{2}\left[1
  + \beta \mu^2 \right]^2 P(k) 
\end{equation} 
\begin{equation}
\bar{T}(z) = 4.0 \text{mK} (1+z)^2 \left(\dfrac{\Omega_b h^2}{0.02}
\right) \left(\dfrac{0.7}{h} \right) \left(\dfrac{H_0}{H(z)} \right)  
\end{equation}
where $P(k)$ is the matter power spectrum at the redshift $z$,
$\bar{x}_{HI}$ is the mean neutral hydrogen fraction.
$\mu = k_{\parallel}/k$ is the cosine of angle between \textbf{k} and the
line of sight and $\beta$ is the linear distortion parameter.
$\beta$ depends on cosmology and bias parameter $b$. 
The $\mu$ dependence of $P_{HI}(\textbf{k})$ arises from the peculiar
velocities and the resulting redshift space distortions
\citep{1987MNRAS.227....1K,1996ApJ...462...25Z}.
We use $\bar{x}_{HI} = 2.45 \times 10^{-2}$ and this corresponds to
${\Omega}_{gas} = 0.001$ \citep{2012A&A...547L...1N, 2013A&A...556A.141Z}.

\begin{figure}
\begin{center}
  \includegraphics[width=5truein]{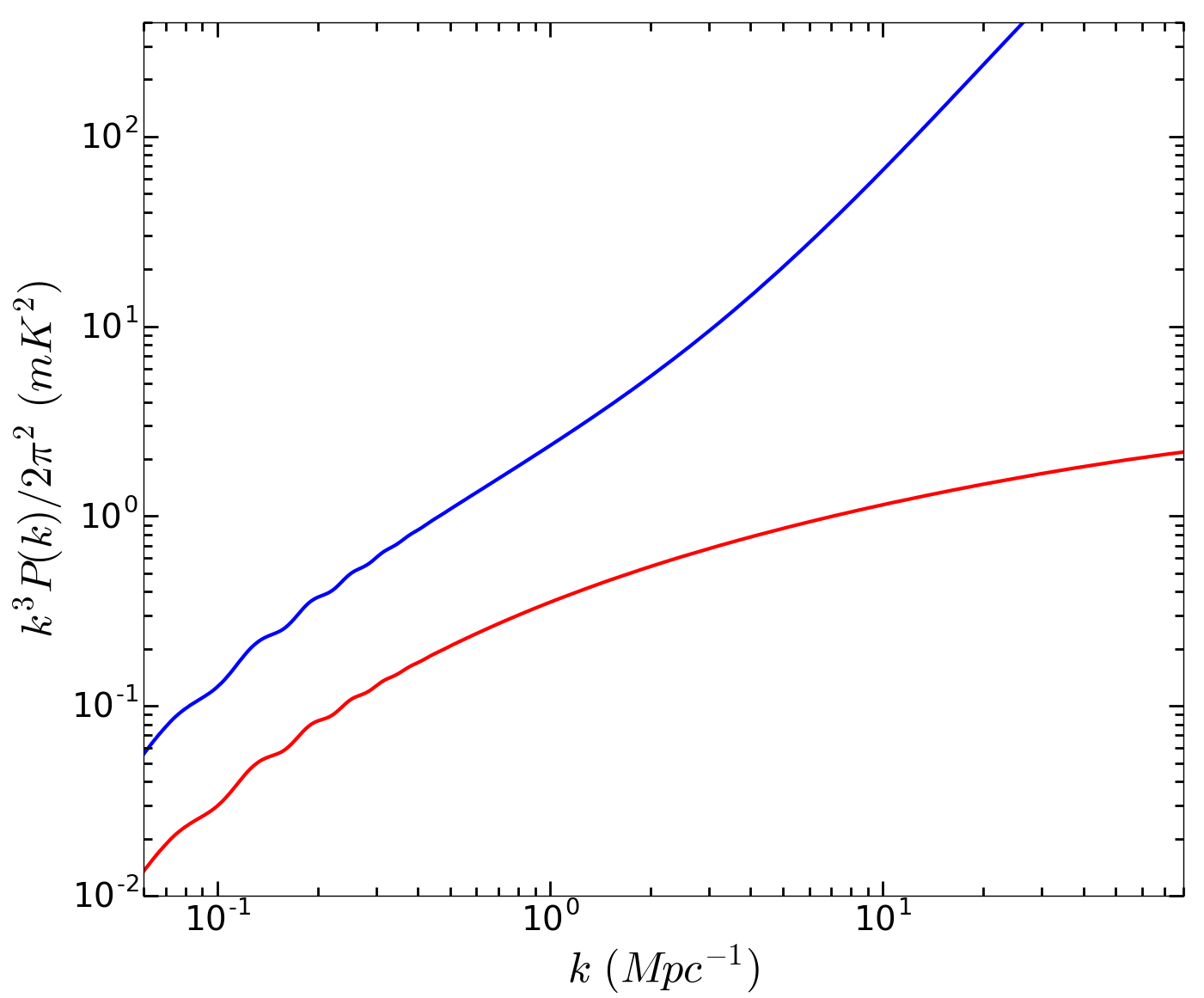}
\end{center}
\caption{This plot shows the dimensionless power spectrum $\Delta^2(k)$
  scaled to brightness temperature units.  The red curve shows linearly
  extrapolated WMAP9 matter power spectrum $P(k)$ for the parameters
  mentioned in the text. The blue curve shows HI power spectrum
  $P_{HI}(k)$ calculated using the bias model mentioned in eq.(7).} 
\end{figure}

N-body simulations \citep{2010MNRAS.407..567B} provide justification
for considering a linear scale independent bias 
$b \simeq 2$ for $k \ll 1 \text{Mpc}^{-1}$ with respect to the linear
power spectrum.
For $k > 1 \text{Mpc}^{-1}$, the bias $b$ has a non-linear dependence
on $k$.
There are two steps involved in the calculation of the 
HI power spectrum: mapping from the linearly extrapolated dark matter
power spectrum to the non-linear power spectrum, and, a scale dependent
bias $b(k)$.
Both of these are obtained from N-Body simulations, but simulations
can access only a finite range of scales.
Limiting our calculations to this range of scales introduces
oscillations in the computed signal hence it is essential to model
these over the entire range of scales.
We choose to do so by fitting a function relating the linearly
evolved power spectrum with the HI power spectrum as obtained in
N-Body simulations.

We have modelled the scale dependence of bias $b\equiv b(k)$ in
which $b(k)$ varies with $k$ as 
\begin{equation}
b(k) = \dfrac{b_0 + b_1(k/k_0)}{1 + b_2(k/k_1)}
\end{equation}
where $b_0 = 2.0, \ b_1/k_0 = 0.6\ \text{Mpc}^{-1}, \ \& \ b_2/k_1 =
0.005\ \text{Mpc}^{-1}$.
The parameters are obtained from the HI power spectrum computed using
an ansatz in N-Body simulations (Model 2 in \citet{2010MNRAS.407..567B}).  
The first two parameters, $b_0$ and $b_1/k_0$ are strongly constrained
by the simulation data.
$b_2/k_1$ is not constrained very strongly and we choose a value that
fits the data and minimises the impact on the estimated signal.  
As mentioned above, this bias is defined with respect to the linearly
evolved dark matter power spectrum and not the non-linear power
spectrum computed from simulations for convenience of calculations.
We have employed this scale dependent bias model (eq. 7) to calculate
the HI power spectrum (eq. 6) which gives a close approximation to the
HI power spectrum calculated using N-body simulations upto $k\sim 10
\ \text{Mpc}^{-1}$.
We have used the linearly extrapolated WMAP 9 matter power spectrum
with $\Lambda$CDM cosmology to calculate HI power spectrum and the
corresponding parameters are $H_0 = 70.0 \text{Km/s/Mpc},
\ {\Omega}_bh^2 = 0.02264, \ {\Omega}_{\Lambda} = 0.721, n_s = 0.972,
  {\sigma}_8 = 0.821$.   

\begin{figure}
\begin{center}
  \includegraphics[width = 5truein]{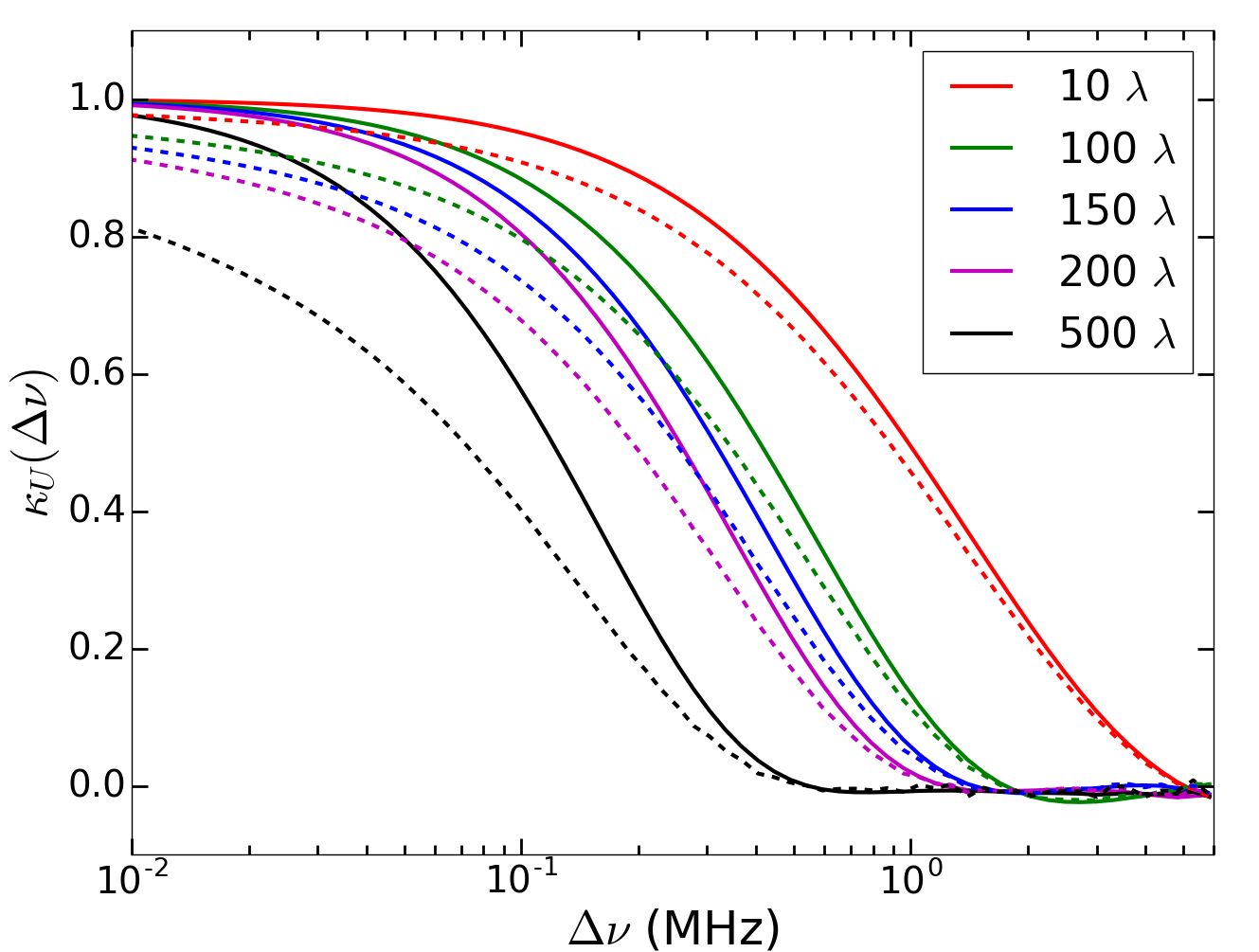}
\end{center}
\caption{Frequency decorrelation function ${\kappa}_{U}(\Delta \nu)$
  as a function of $\Delta \nu$ at five different \textit{U} values
  indicated in figure. The signal decorrelates more sharply for higher
  value of \textit{U}.  Dashed lines show the decorrelation function
  for the model with scale dependent bias whereas solid lines show the
  corresponding function for linear bias.  It is seen here that the
  signal remains correlated over a smaller range in frequency for the
  scale dependent bias model as compared to the linear bias model.  This
  indicates that including the effect of scale dependent bias decreases the
  frequency correlation of signal.  This decrement is more
  significant at higher values of \textit{U}.} 
\end{figure}

We use the decorrelation function ${\kappa}_{U}(\Delta \nu)$ to quantify
the $\Delta \nu$ dependence of the signal $S_2(U, \Delta \nu)$.
This allows us to assess the HI signal decorrelation with frequency
separation $\Delta \nu$.
${\kappa}_{U}(\Delta
\nu)$ can be defined as \citet{2007MNRAS.378..119D}
\begin{equation}
{\kappa}_{U}(\Delta \nu) = \dfrac{S_2(U, \Delta \nu)}{S_2(U, 0)}
\end{equation}

Figure 2 shows ${\kappa}_{U}(\Delta \nu)$ as a function of $\Delta
\nu$ for different baselines \textit{U}.
Dashed lines show the variation for the scale dependent bias and solid
lines for the constant bias as mentioned above.
In both models the bias is applied to the linearly extrapolated power
spectrum as this facilitates comparison with results published in
\citet{2015JApA...36..385B}. 
The signal is fully correlated at zero separation, as expected, and we
have ${\kappa}_{U}(0) = 1$.
The correlation falls (${\kappa}_{U}(\Delta
\nu) < 1$) as $\Delta \nu$ is increased.
We see that ${\kappa}_{U}(\Delta \nu)$ varies slowly with $\Delta \nu$
at the small baselines.
For $U = 10$, we have ${\kappa}_{U}(\Delta \nu) = 0.5$ at
$\Delta \nu \approx 1 \text{MHz}$, beyond which $\kappa$ falls
further.
At larger baselines, ${\kappa}_{U}(\Delta \nu)$ has a steeper
dependence with $\Delta \nu$.
For $U = 200$, ${\kappa}_{U}(\Delta \nu) =
0.5$ at $\Delta \nu \approx 0.2 \text{MHz}$, and ${\kappa}_{U}$
crosses zero near $\Delta \nu \approx 1 \text{MHz}$.
The signal decorrelates faster for the scale dependent bias model
where there is more power at small scales. 
This plot illustrates that the combined boost at small scales due to
non-linear evolution and scale dependent bias is
counter-balanced by the smaller frequency range over which the
signal is correlated.

Following \citep{2014JApA...35..157A}, we define $\Delta {\nu}_{0.9}$,
$\Delta {\nu}_{0.5}$ and $\Delta 
{\nu}_{0.1}$ as the values of the frequency separation $\Delta {\nu}$
where the decorrelation falls to $0.9$, $0.5$ and $0.1$ respectively
i.e. ${\kappa}_{U}(\Delta {\nu}_{0.5}) = 0.5$, etc.
$\Delta {\nu}_{0.9}$, $\Delta {\nu}_{0.5}$ and $\Delta
{\nu}_{0.1}$ are shown as a function of \textit{U} in Figure~3.
The oscillations visible in ($\Delta \nu,U$) space on the contours of
${\kappa}_{U}(\Delta \nu) = 0.1$ and $0.5$ represent the BAO (Baryon
Acoustic Oscillations) feature present in the power spectrum $P(k)$.
As in Figure~2, the dashed lines are for the model with scale dependent
bias and solid lines are for constant bias model.  

\begin{figure}
\begin{center}
  \includegraphics[width = 5truein]{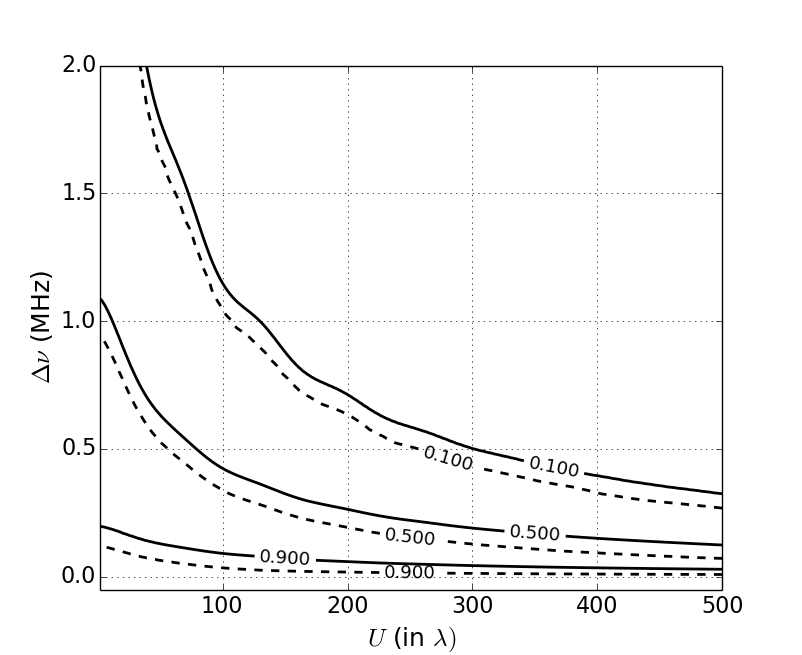}
\end{center}
\caption{This figure shows how $\Delta \nu$ is varying as a function
  of \textit{U} for a given values of frequency decorrelation function
  ${\kappa}_{U}(\Delta \nu)$ indicated in the figure.  Dashed lines
  show the relation for the scale dependent bias model whereas solid
  lines are for the linear bias model.  As seen in figure~2, we find
  that the signal remains correlated to larger frequency differences
  with linear bias.  The baryon acoustic oscillations leaves a
  clear signature in the decorrelation.} 
\end{figure}

\begin{figure}
\begin{center}
  \includegraphics[width=7truein]{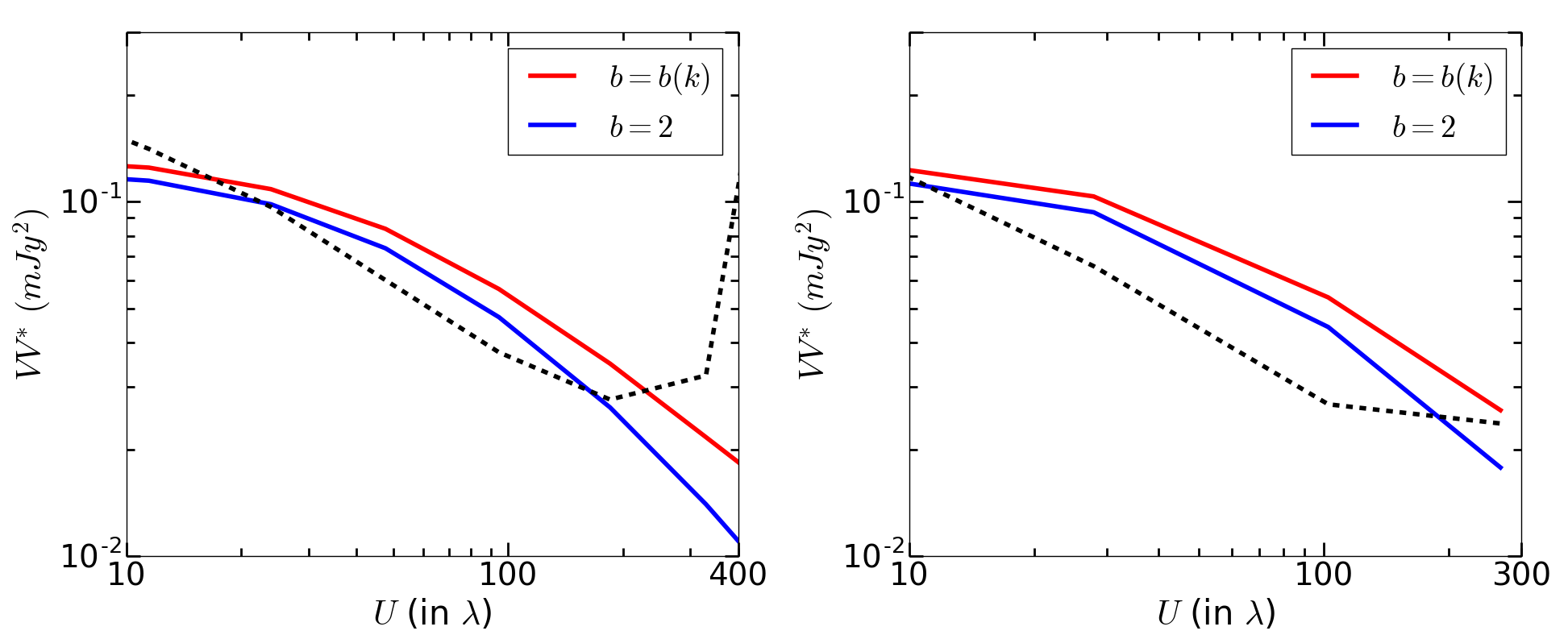}
\end{center}
\caption{The expected visibility correlation $VV^* = S_2(U,\Delta \nu)$ as a
  function of \textit{U} for OWFA.  The solid (red) curve shows the 
  expected signal and dashed (black) curve shows the expected noise in
  the visibility correlation. The solid blue line is for scale
  independent bias whereas the red line is for scale dependent bias +
  non-linear evolution of perturbation.  In both the plots the red
  curve sits above the blue curve by $10\% - 30\%$ showing that the
  scale dependent effects are important for calculation of the
  signal.  The noise is calculated for 1000 Hrs of
  integration. This is shown for signal binned in 9 (left panel)and 5
  (right panel)logarithmic bins. We observe that $1\sigma$ detection is
  possible with $10^3$ hrs of integration for the signal in baseline
  range of $U \sim 20 - 200$.} 
\end{figure}

\begin{figure}
\begin{center}
  \includegraphics[width=7truein]{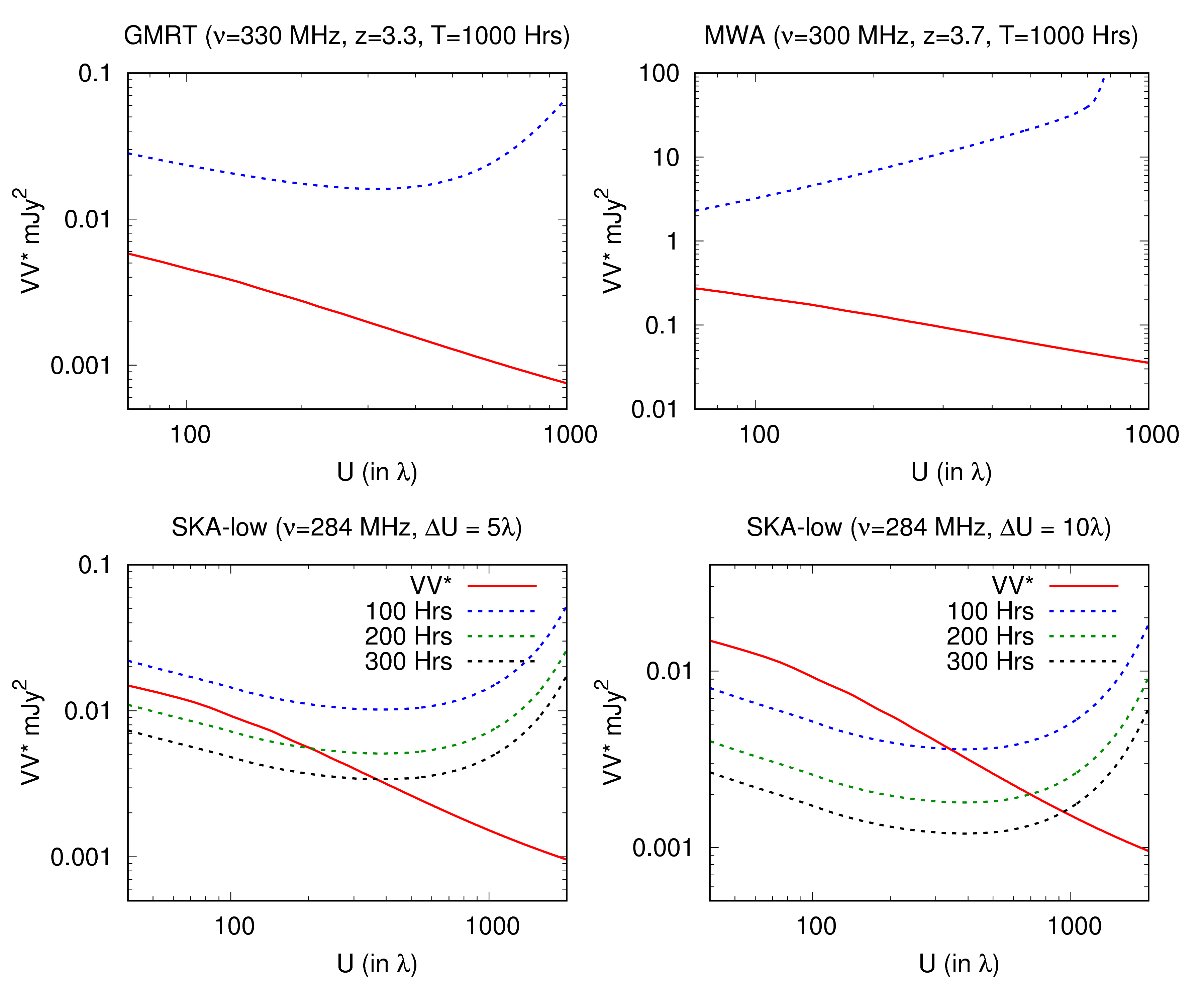}
\end{center}
\caption{Expected visibility correlation $VV^* = S_2(U,\Delta \nu)$ as a
  function of \textit{U}. Solid (red) curve shows expected signal and dashed
  (blue) curve shows expected noise in visibility correlation. For
  GMRT and MWA, noise is calculated for 1000 Hrs of integration. Top
  left panel shows the expected signal for GMRT. Top right panel shows
  expected signal for MWA. Left bottom panel shows expected signal for
  SKA-low ($\nu = 284$ MHz) corresponding to bin size $\Delta U =
  5\lambda$ for different integration times. Right bottom panel shows
  expected signal for SKA-low corresponding to the bin size $\Delta U =
  10\lambda$.}  
\end{figure}

\section{Noise in visibility correlations}

We have used the HI power spectrum $P_{HI}(k)$ shown in Figure~1 (blue
curve) to compute visibility correlations using the approach described
above. 
We have also estimated the noise level for OWFA and compared it with
the expected signal.   

The real part of ${\cal N}(U_n,\nu)$ has {\sl rms} fluctuation
$\sigma$ given by 
\begin{equation}
\sigma = \dfrac{\sqrt{2}k_B T_{sys}}{\eta A \sqrt{\Delta {\nu}_c \Delta t}}
\end{equation}
Values of $T_{sys}$, $\eta$ and $A$ used in our calculations are given
in Table 1. 

As discussed in \citep{2014JApA...35..157A}, it is possible to avoid
the noise contribution ${N}_{2}(U_n,\Delta \nu)$ in the visibility
correlation ${V}_{2}(U_n,\Delta \nu)$ by correlating only those
visibility measurements where the noise is uncorrelated.
In this case, for a fixed baseline \textit{U} we only correlate the
visibilities measured by different redundant antenna pairs or the
visibilities measured at different time instants. 
OWFA has a large redundancy in baselines and redundant baselines
provide many independent estimates of the visibility correlation
$(\Delta V_{2})$ at the same \textit{U} where each estimate has an
independent system noise contribution, but signal is the
same.
The OWFA is equatorially mounted, hence the projected baseline
lengths do not change as the source is tracked.
The noise in the visibilities measured with different antenna pairs is
uncorrelated.  
The noise in the visibilities measured at two different time instants is
also uncorrelated.
Given that the signal decorrelates within $\Delta\nu \simeq 1 MHz$,
whereas the bandwidth is much larger, the observing bandwidth $B$ also
provides several independent estimates of the visibility correlation.
Taking all of these into consideration, we get:
\begin{equation}
(\Delta N_{2})^{2} = \left(  \dfrac{2{\sigma}^{2} \Delta
    t}{t_{obs}}\right)^{2} \dfrac{\Delta {\nu}_{0.5}}{N_{P}B}
  \ \text{and} \ (\Delta S_{2})^{2} = \dfrac{(S_2)^2\Delta
    {\nu}_{0.5}}{N_{E}B} 
\end{equation}
where $\Delta S_{2}$ and $\Delta N_{2}$ are cosmic variance and system
noise contribution respectively.
$\Delta t$ is the the correlator integration time and $t_{obs}$ is the
total observation time.
$N_P$ and $N_E$ denote the number of independent estimates of the
system noise and the signal respectively in each baseline bin.
$\Delta {\nu}_{0.5}$ is the frequency at which signal decorrelates to
$0.5$ times the value at $\Delta {\nu} = 0$ and $B$ is the frequency
bandwidth of the instrument.
For more details about the noise calculation please see
\citet{2005MNRAS.356.1519B, 2007MNRAS.378..119D, 2010MNRAS.407..567B}.

The total error $\Delta V_{2}$ in the residual visibility is 
\begin{equation}
\sqrt{(\Delta V_{2})^{2}} = \sqrt{(\Delta S_{2})^{2} + (\Delta N_{2})^{2}}
\end{equation}

Detection is possible with $10^3$ hrs of integration for the signal in
baseline range of $U \sim 20 - 200$ (Figure~4).
We can optimize bin size to improve prospects for detection.  
We see that detection with significance $2\sigma$ is
possible with $10^3$ hrs of integration for the signal binned with $5$
logarithmic bins on baseline scale for the baseline range of
$U \sim 20 - 200$ (Figure~4).  

We have done a similar analysis for
GMRT\footnote{http://www.gmrt.ncra.tifr.res.in},
MWA\footnote{http://www.mwatelescope.org/} and upcoming
SKA-low\footnote{https://www.skatelescope.org/} to compare the results
for OWFA with these instruments.   
Figure~5 shows the expected signal and system noise calculated for
GMRT, MWA and SKA-low respectively. 
We find that HI detection is not possible with either GMRT or MWA for
$10^3$ hrs of integration time without aggressive binning whereas
SKA-low can detect HI signal within 200 Hrs of integration (Figure~5
bottom panels).  

We have investigated the Signal to Noise Ratio (SNR) for detecting the
HI signal under the assumption that it is possible to completely
remove the foregrounds.
Figure~6 shows the SNR for OWFA as a function of baselines with data
binned in $9$ and $5$ logarithmic bins respectively.
We see that $>2\sigma$ detection of the signal is possible in the
baseline range $20\leq U \leq 200$ in $\sim 2000$ hrs of
integration (Figure~6) for data divided in $9$ logarithmic bins and,
for $5$ logarithmic bins, $>3\sigma$ detection of the signal is
possible in the baseline range $30\leq U \leq 150$ in $\sim
2000$ hrs of integration (Figure~6) and $>5\sigma$ detection in
baseline range $50\leq U \leq 130$ in $\sim 3000$ hrs of
integration.
We observe that optimizing bin size enhances Signal to Noise Ratio
which in turn leads to signal detection with better significance
levels. 
Calculations done using the Fisher matrix approach indicate that
a $5\sigma$ detection of the HI power spectrum via measurement of the 
amplitude of the HI power spectrum is possible in $150$~hours
\citep{2015JApA...36..385B}. 
In a later work they have shown \citep{OWFA.HIPK} that
a measurement of the binned HI power spectrum in the relevant range of
scales is possible in $10^3$~hours. 
Thus it is possible to optimize combining signal in different modes
well beyond naive binning that we have considered here.

\begin{figure}
  \begin{center}
    \includegraphics[width=7truein]{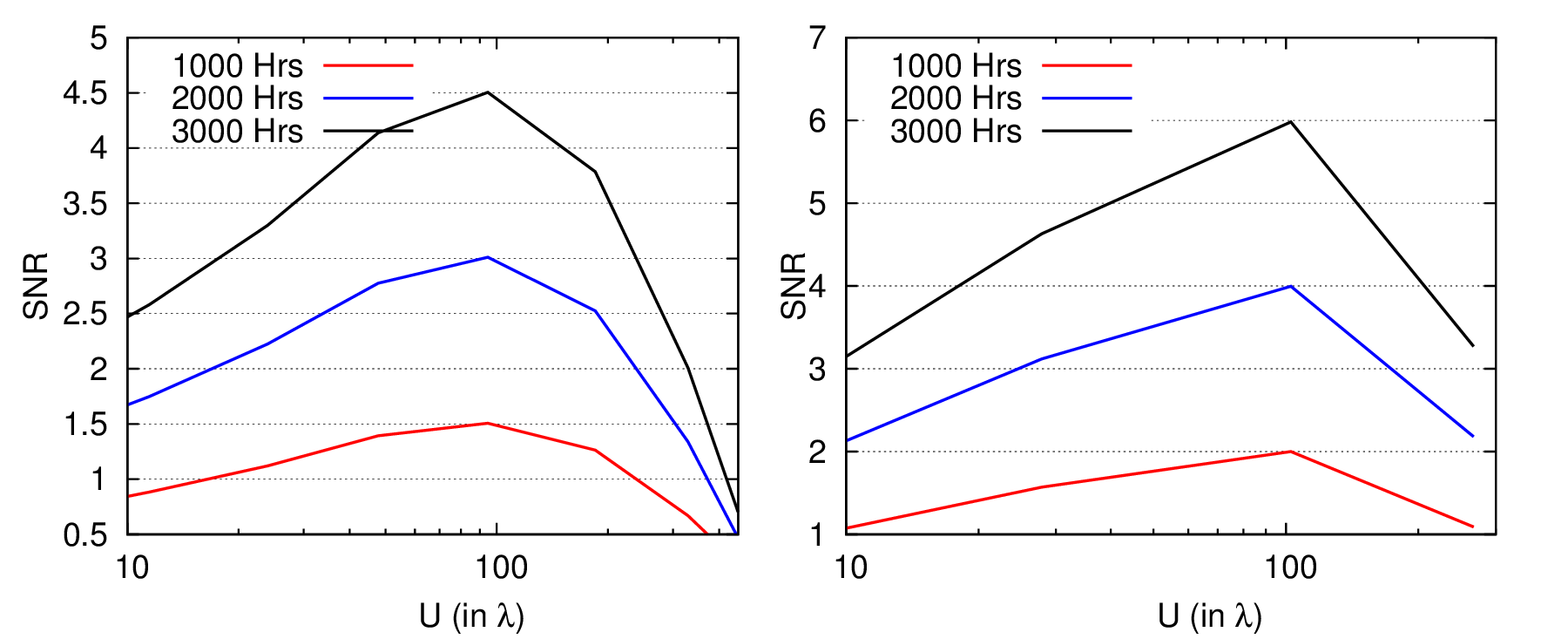}
  \end{center}
\caption{This figure shows the signal to noise ratio (SNR) as function
  of baseline \textbf{U} with data divided in $9$ logarithmic bins (left
  panel) and $5$ logarithmic bins (right panel) for different
  integration times indicated.}  
\end{figure}

\section{Summary}

We have compared HI detection prospects using upgraded OWFA with GMRT,
MWA and SKA-low.
OWFA can detect HI signal in about $2000$ Hrs of integration. 
HI signal detection is not possible with GMRT ($330$~MHz) and MWA
($300$~MHz) for a similar integration.
SKA-low ($284$~MHz) can detect HI signal within $200$ Hrs of
integration.
From the investigation of SNR with different $U$ bin we observe that
optimization of $U$ bins can enhance SNR.  
Significant optimization may be possible using the Fisher matrix
approach, and that will be applicable in a similar manner to all
instruments apart from subtle variations depending on the relative
role of cosmic variance. 

We have introduced the concept of a functional form of scale dependent
bias defined with respect to the linearly extrapolated dark matter
power spectrum.
This combines the effects of non-linear evolution and scale dependent
bias. 
This is a useful tool as it allows us to calculate quantities without
limiting us to the range of wave modes available in the N-Body
simulations that are used to compute bias. 

The non-linear evolution of power spectrum and the scale dependent
bias enhance the HI power spectrum, but we also see that the signal
decorrelates within a shorter frequency range when these factors are
taken into account.
It is to be noted that we have taken only linear mapping in red-shift
space and a direct estimation of the decorrelation function from
N-Body simulations is required for a clearer picture.

The prospects for detection of HI distribution at $z \sim 3$ with OWFA
are encouraging.  
Optimizing the bin sizes and using multiple realizations can lead to
detection with higher significance level in smaller integration
times. 
Indeed, Fisher matrix based analysis strongly suggests that
detection of the amplitude of power spectrum may be possible in a
couple of hundred hours while determination of the shape of the power
spectrum will require close to $10^3$ hours
\citep{2015JApA...36..385B, OWFA.HIPK}.
The comparative analysis suggests that at present OWFA
is perhaps the most promising instrument at present for this redshift
window, and will continue to be up to the time when SKA-low becomes
operational. 
Given the projected time line for SKA, it is clear that OWFA can make
significant contribution in the interim through detection of HI at $z
\sim 3$.

In our discussion, we have ignored the role of foregrounds and foreground
subtraction.  This is likely to impact prospects of detection of the
redshifted $21$~cm radiation from intermediate and high redshifts.

The upgraded GMRT (uGMRT) with its wide band receivers will be
available soon.
This enables observations of a larger comoving volume as compared to
the GMRT in a single observation.
The expected improvement in sensitivity as a result of wider band is
about a factor $3$, and hence with some imaginative co-adding of
signal it should be possible to detect the signal in $10^3$~hours or
so.
The uGMRT can play an important role in detection of redshifted
$21$~cm radiation from $z \sim 3$.
Predictions for uGMRT need to account for the evolution of signal as
the wide band encompasses a large range of redshifts.

\section*{Acknowledgments}

Computational work for this study was carried out at the cluster
computing facility in the IISER Mohali.    
This research has made use of NASA's Astrophysics Data System. 
The authors would like to thank Jayaram Chengalur, Saiyad Ali and
Somnath Bharadwaj for useful comments and discussions.

\end{document}